\documentclass[10pt,reprint,prb,showpacs,superscriptaddress,floatfix,amsmath,amssymb,amsfonts,aps]{revtex4-1}
\usepackage{graphicx}
\usepackage{multirow}
\newcommand{\cupo}{$\alpha$-Cu$_2$P$_2$O$_7$}
\newcommand{\cuas}{$\alpha$-Cu$_2$As$_2$O$_7$}
\newcommand{\cuvo}{$\beta$-Cu$_2$V$_2$O$_7$}
\newcommand{\cua}{Cu$_2A_2$O$_7$}
\newcommand{\cuaf}{Cu$_2A_2$O$_7$ ($A$\,=\,P,\,As,\,V)}
\begin{document}

\title{Peculiar long-range superexchange in Cu$_2A_2$O$_7$ ($A$ = P, As,
V)\\ as a key element of the microscopic magnetic model}

\author{O. Janson}
\email{janson@cpfs.mpg.de}
\affiliation{Max-Planck-Institut f\"{u}r Chemische Physik fester
Stoffe, 01187 Dresden, Germany}

\author{A.~A. Tsirlin}
\email{altsirlin@gmail.com}
\affiliation{Max-Planck-Institut f\"{u}r Chemische Physik fester
Stoffe, 01187 Dresden, Germany}

\author{J. Sichelschmidt}
\affiliation{Max-Planck-Institut f\"{u}r Chemische Physik fester
Stoffe, 01187 Dresden, Germany}

\author{Y. Skourski}
\affiliation{Hochfeld-Magnetlabor Dresden, Helmholtz-Zentrum
Dresden-Rossendorf, 01314 Dresden, Germany}

\author{F. Weickert}
\email{Present address: LANL, Los Alamos, New Mexico 87545, USA}
\affiliation{Max-Planck-Institut f\"{u}r Chemische Physik fester
Stoffe, 01187 Dresden, Germany}
\affiliation{Hochfeld-Magnetlabor Dresden, Helmholtz-Zentrum
Dresden-Rossendorf, 01314 Dresden, Germany}

\author{H. Rosner}
\email{rosner@cpfs.mpg.de}
\affiliation{Max-Planck-Institut f\"{u}r Chemische Physik fester
Stoffe, 01187 Dresden, Germany}

\date{\today}

\begin{abstract}
A microscopic  magnetic model for $\alpha$-Cu$_2$P$_2$O$_7$ is evaluated in a
combined theoretical and experimental study. Despite a dominant intradimer
coupling $J_1$, sizable interdimer couplings enforce long-range magnetic
ordering at $T_N$\,=\,27\,K.  The spin model for $\alpha$-Cu$_2$P$_2$O$_7$
is compared to the models of the isostructural $\beta$-Cu$_2$V$_2$O$_7$
and $\alpha$-Cu$_2$As$_2$O$_7$ systems.  As a surprise, coupled dimers
in $\alpha$-Cu$_2$P$_2$O$_7$ and alternating chains in
$\alpha$-Cu$_2$As$_2$O$_7$ contrast with a honeycomb lattice in
$\beta$-Cu$_2$V$_2$O$_7$.  We find that the qualitative difference in
the coupling regime of these isostructural compounds is governed by the
nature of $A$O$_4$ side groups: $d$-elements ($A$\,=\,V) hybridize with
nearby O atoms forming a Cu--O--$A$--O--Cu superexchange path, while for
$p$-elements ($A$\,=\,P,\,As) the superexchange is realized via O--O edges of
the tetrahedron. Implications for a broad range of systems are
discussed.
\end{abstract}


\maketitle

\section{Introduction}
The search for new magnetic ground states (GS) is an arduous
challenge for condensed matter physics. Since about two decades ago,
the rapid development of computational techniques has spurred theoretical
studies on various magnetic models. The recent observation of
magnetic monopoles,\cite{magn_monopol_exp} anticipated by a
theoretical prediction,\cite{magn_monopol_theor} is
one of the most remarkable examples evidencing the excellent
potential and reliability of sound theoretical studies.

Still, it is often tricky to find the way from a theoretical result
to its experimental confirmation. In particular, the applicability of a
theoretical model to a real material should be addressed.  For
magnetic systems, inelastic neutron scattering (INS) experiments,
which yield momentum-resolved excitation spectra for a wide range of
energies and temperatures, are typically a method of choice.
However, even the powerful neutron techniques are insecure
against the ambiguous interpretation of experimental results. Thus, for
instance, the
analysis of INS data in Ref.~\onlinecite{AHC_VO2P2O7_INS_chain_plus}
correctly disclosed the irrelevance of the two-leg Heisenberg
spin-ladder model for the ambient-pressure modification of
(VO)$_2$P$_2$O$_7$. At the same time, the experimental spectra 
exhibited a pronounced high-energy mode, and the authors of
Ref.~\onlinecite{AHC_VO2P2O7_INS_chain_plus} interpreted it as a
signature of a two-magnon bound state. Subsequent studies based on
alternative techniques refuted this conjecture and revealed that the
high-energy mode and the double-gap feature arise from two
inequivalent alternating Heisenberg chains (AHC) present in this
system.\cite{AHC_VO2P2O7_two_chains_MH_NMR} As another example, the
frustrated Heisenberg chain LiCu$_2$O$_2$, had been proposed to imply
two relevant antiferromagnetic (AF) exchange integrals $J_1$ and $J_2$,
based on thermodynamical measurements and neutron diffraction
data,\cite{FHC_LiCu2O2_chiT_CpT_NS_wrong_model_paper,
*FHC_LiCu2O2_chiT_CpT_NS_wrong_model_comment,
*FHC_LiCu2O2_chiT_CpT_NS_wrong_model_paper_reply} while the combined
nuclear magnetic resonance (NMR) and density functional theory (DFT)
study (Ref.~\onlinecite{FHC_LiCu2O2_NMR_DFT}) as well as the INS investigation
(Ref.~\onlinecite{FHC_LiCu2O2_INS}) revealed the ferromagnetic (FM)
nature of $J_1$.

\begin{figure}[tbp]
\includegraphics[width=8.6cm]{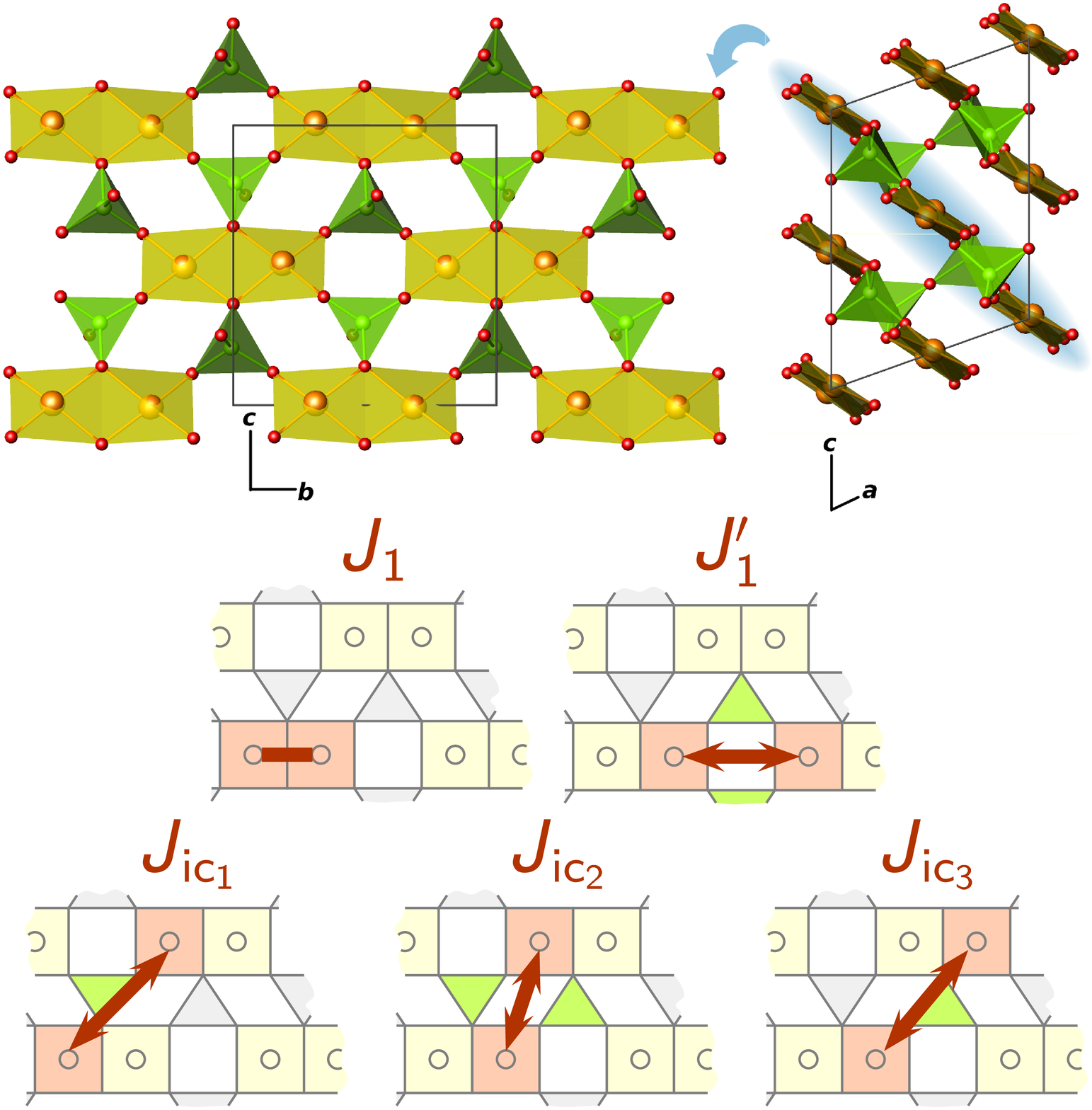}
\caption{\label{F_str}(Color online) Crystal structure and exchange
coupling paths in \cuaf. Squares and
triangles represent CuO$_4$ plaquettes and $A$O$_4$ tetrahedra,
respectively.} 
\end{figure}

Recently, density functional theory (DFT) calculations developed into an
appealing alternative, complementary to experimental techniques.  At
present, the performance and reliability of DFT codes allow us to evaluate
quantitative magnetic models even for complex materials, having only
crystallographic information at hand. Moreover, DFT calculations are
capable of even improving the structural input,\cite{CuClLaNb2O7_DFT,
*CuXLaNb2O7_DFT_str, *cucl-lanb-structure,*cucl-lanb-structure-2} thus
overcoming the problem of weakly scattering elements (for instance, H
for X-rays or V for neutrons) and the subsequently large error bars in
the experimental atomic positions.

The inverse problem --- to find a crystal structure for a specific spin
lattice --- is even more complex. Here, semiempirical
Goodenough-Kanamori-Anderson (GKA) rules,\cite{GKA_1, *GKA_2, *GKA_3}
formulated about 60 years ago, still provide invaluable support.
However, these rules are at best not universal. In particular, the GKA
rules cannot be applied to long-range couplings that involve more than
one ligand atoms. The abundance of such couplings in real materials
strongly motivates the development of further rules to account for these
situations.

Semiempirical rules are especially helpful for a fine tuning of the
material, because chemical substitutions usually alter exchange
couplings or even the spin lattice. For example, the substitution of Br
for Cl changes the singlet ground state of (CuCl)LaNb$_2$O$_7$ toward
columnar AF ordering in the bromine compound.\cite{cubr-lanb} The
replacement of Zn by Mg in kapellasite Cu$_3$Zn(OH)$_6$Cl$_2$ leads to
haydeeite Cu$_3$Mg(OH)$_6$Cl$_2$ with a reduced nearest-neighbor (NN) coupling and
modified spin physics.\cite{kapel_hayd_DFT}

In the present study, we perform a comparative investigation of the
isostructural \cuaf~compounds and consider an instructive example that 
is relevant for a family of systems with tetrahedral anions. The
pentavalent nature of the $A$ atoms ensures the magnetic $3d^9$
electronic configuration of Cu$^{2+}$. The thortveitite-type crystal structure of \cua~(Fig.~\ref{F_str}, top)
contains magnetic layers of Cu$_2$O$_6$ structural dimers (two
edge-sharing CuO$_4$ plaquettes) coupled by non-magnetic
$A$O$_4$ tetrahedra. The neighboring layers are connected by
two $A$O$_4$ tetrahedra forming $A_2$O$_7$ pyrophosphate, pyroarsenate
or pyrovanadate groups (Fig.~\ref{F_str}, top right).

In this study, we disclose the underlying reason for a variety
of magnetic behaviors --- coupled spin dimers in \cupo, alternating spin
chains in \cuas, and a honeycomb spin lattice in \cuvo~--- realized
within the same basic crystal structure. Using DFT calculations, we
evaluate five relevant exchange couplings (Fig.~\ref{F_str}, bottom).
Surprisingly, these couplings are drastically affected by the electronic
state of the side groups' central atom $A$, whether it is a $p$ (P, As)
or a $d$ (V) element.  Understanding the crucial role of non-magnetic
$A$O$_4$ groups is the key for directed chemical substitution, which may
open the door for a tunable magnetism.

The magnetic properties of \cuvo~have been actively studied.\cite{*[{}] [{,
and references therein.}] HCL_Cu2V2O7_DFT_GGA_AHC_model} Very recently,
we derived the actual spin model and identified an almost ideal spin-1/2
honeycomb lattice in this compound.\cite{HCL_Cu2V2O7_DFT_DCC_simul}
Considering the crystal structure of \cuvo, this magnetic model looks
highly non-trivial.  Puzzled by the driving force of such an unusual
magnetic coupling regime, we prepared the isostructural system \cupo~and
investigated its magnetic behavior to allow for a comparative study.

Cu$_2$P$_2$O$_7$ has two polymorphic modifications. The low-temperature
($\alpha$) and the high-temperature ($\beta$) phases reveal similar
monoclinic crystal structures (Fig.\,\ref{F_str}), yet the
$\beta$-modification shows a twice smaller $c$ lattice parameter and
disordered arrangement of the oxygen atom shared by the two PO$_4$
tetrahedra of the pyrophosphate group.\cite{structure2006} In the following, we consider the
$\alpha$-modification which is fully ordered and stable below 350\,K.  We
would like to emphasize that this $\alpha$-modification is isostructural to
the $\beta$-modification of Cu$_2$V$_2$O$_7$. The magnetic behavior of
\cupo\ was studied in the early 1970s. An AF ordering was found at 27\,K, with
tentative models of the magnetic structure established from neutron
diffraction\cite{neutrons} and NMR.\cite{nmr} However, no information on the microscopic magnetic
model has been reported.

This paper is organized as follows. We introduce the methods used in
our study in Sec.\,\ref{methods}. In Sec.\,\ref{cupo}, we present
our experimental as well as theoretical investigation of \cupo.
In Sec.\,\ref{disc}, we compare the magnetic models of \cuaf, and
discuss the origin of their essentially different nature. Finally,
in Sec.\,\ref{summ}, we summarize the results and give a short
outlook.

\section{\label{methods}Methods}
A single-phase powder sample of \cupo\ was prepared by a
solid-state reaction of CuO and NH$_4$H$_2$PO$_4$. A stoichiometric
mixture of the reagents was fired at 250\,$^{\circ}$C for 6\,h in air to
remove ammonia and water. The sample was further placed into a sealed
quartz tube and heated at 800\,$^{\circ}$C for 12\,h. The resulting light-gray
powder was analyzed by X-ray diffraction (Huber G670 Guinier camera,
Cu$K_{\alpha1}$ radiation, image plate detector, 3--100$^{\circ}$ angle
range). A Rietveld refinement confirmed the formation of \cupo\ and did
not show any traces of impurity phases (in particular, the admixture of
$\beta$-Cu$_2$P$_2$O$_7$ can be ruled out). In contrast, a 
high-temperature annealing in air always produced a Cu$_3$(PO$_4)_2$
impurity.

Magnetic susceptibility ($\chi$) was measured with a SQUID magnetometer (Quantum
Design MPMS) in the temperature range 2--380\,K in applied
fields up to 5\,T. High-field magnetization measurements (at $T$\,=\,1.4\,K) were
performed at the Dresden High Magnetic Field Laboratory in pulsed fields
up to 60\,T.  Details of the experimental procedure are described
elsewhere.\cite{tsirlin2009}

The electron spin resonance (ESR) spectra of a powder sample of \cupo\ were recorded at $X$-band
frequencies (9.4 GHz) for temperatures 5--300\,K.  The spectra were
fitted with a powder average of two Lorentzian lines corresponding to
two (effective) components of the $g$-tensor in axial symmetry:
$g_\perp$ and $g_\|$.

DFT calculations were performed using the full-potential code
\texttt{fplo9.00-31}.\cite{FPLO} Within the local (spin) density
approximation (L(S)DA), the exchange and correlation potential of Perdew
and Wang has been used.\cite{PW}

For
the LDA calculations, we chose a mesh of 1008 $k$ points (207 in the
irreducible wedge).
Spin-polarized supercell LSDA+$U$ calculations were performed on 12,
18, and 32 $k$ points meshes.\footnote{Supercells contain a doubled
number of atoms compared to the crystallographic unit cell. In
addition, spin-polarized calculations require a quadrupled matrix
compared to LDA calculations. This leads to a substantial increase
in the computation time. Therefore, only less dense $k$-meshes
can be used.} 
All calculations were done for the experimental crystal structure
from Ref.~\onlinecite{structure2006}. To address the consistency of the
structural data, we calculated forces and cross-checked the LDA results by
using the generalized gradient approximation (GGA). Since the resulting
forces do not exceed 1\,eV/\AA\ $[$the mean value is 0.7(3)\,eV/\AA\ in both
LDA and GGA$]$, no further relaxation has been performed.  The
calculations were carefully checked for convergence.

Quantum Monte-Carlo (QMC) simulations were performed on 24$\times$24
finite lattices with periodic boundary conditions using the
\texttt{looper}\cite{looper} and \texttt{dirloop\_sse} (directed loop in the
stochastic series expansion representation)\cite{dirloop} algorithms from the
\texttt{ALPS} package.\cite{ALPS}

\section{\label{cupo}Magnetism of \cupo}
\subsection{Magnetization and ESR measurements}

\begin{figure}[tbp]
\includegraphics[width=8.6cm]{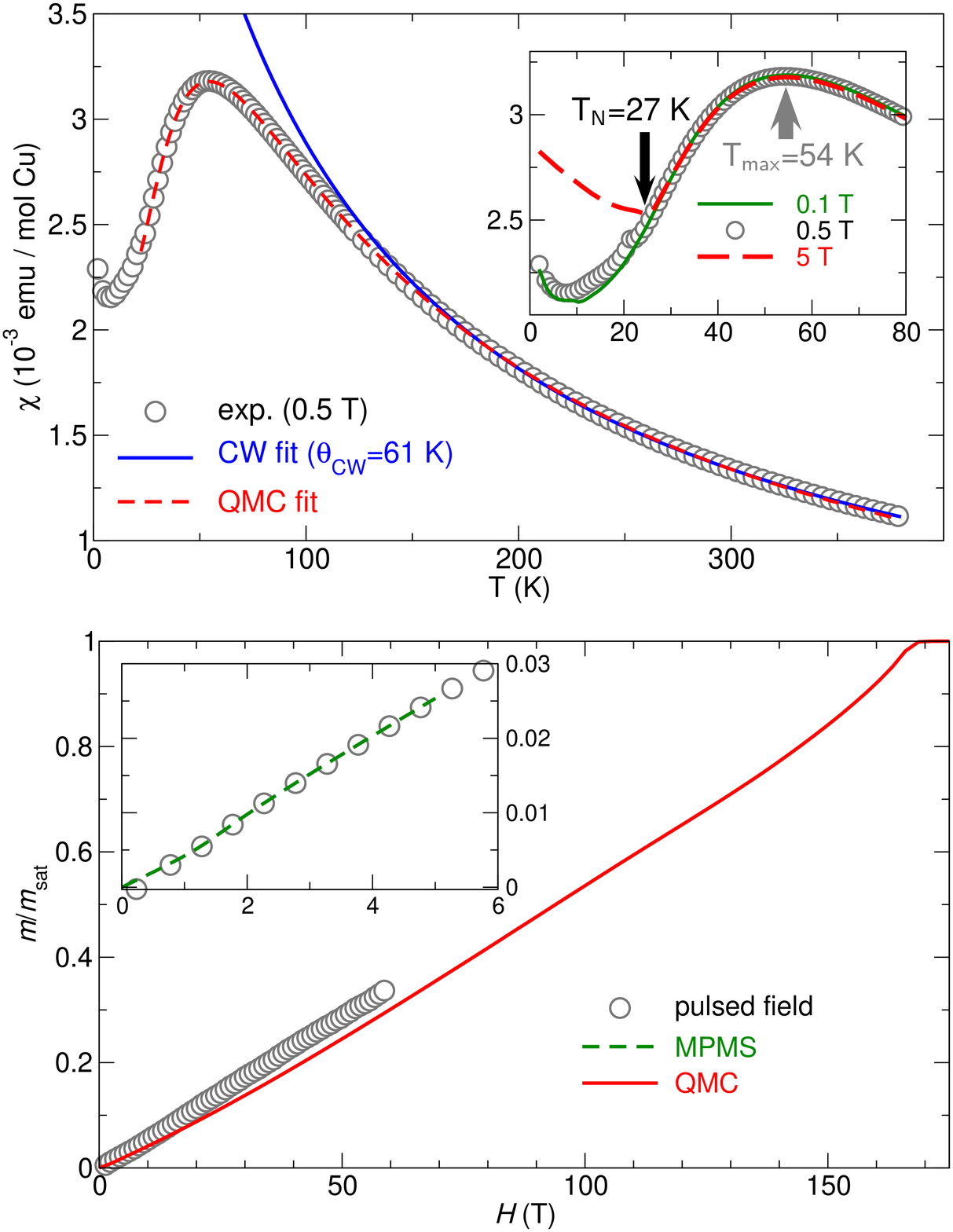}
\caption{\label{F_exp}(Color online) Top: magnetic susceptibility
$\chi(T)$ of Cu$_2$P$_2$O$_7$ together with the Curie-Weiss (CW)
and the quantum Monte-Carlo (QMC) fits (see text). Inset: field
dependence of $\chi(T)$ Bottom: high-field magnetization of
Cu$_2$P$_2$O$_7$ with a QMC fit. Inset:
scaling of the high-field magnetization curve with the MPMS data.}
\end{figure}

The magnetic susceptibility $\chi(T)$ (Fig.~\ref{F_exp}, top) exhibits a
broad maximum at $T_{\text{max}}$\,=\,54\,K indicative of low-dimensional
behavior. The Curie-Weiss (CW) fit yields an AF Weiss temperature
$\theta_{\text{CW}}$\,=\,61\,K and the effective moment of $1.89$\,$\mu_B$,
slightly larger than the spin-only effective moment
$\mu_{\text{eff}}\,=\,2\sqrt{S(S+1)}$\,$\approx$\,1.73\,$\mu_B$ for $S$\,=\,1/2. The deviation from
the CW fit below $\sim140$\,K originates from the onset of AF correlations
that lead to a long-range magnetic ordering at low temperatures.

The magnetic ordering transition is evidenced by a weak kink in the low-field
$\chi(T)$ curves around $T_N\simeq\,$27\,K (circles and solid line in the upper
inset of Fig.~\ref{F_exp}). The data measured at 5\,T reveal a well-defined kink
at the same temperature. The difference between the data collected in low and
high fields below $T_N$ should be related to the spin-flop transition
around 1\,T. The susceptibility upturn below 10\,K is likely caused by a
paramagnetic impurity contribution.

\begin{figure}[tbp]
\includegraphics[width=8.6cm]{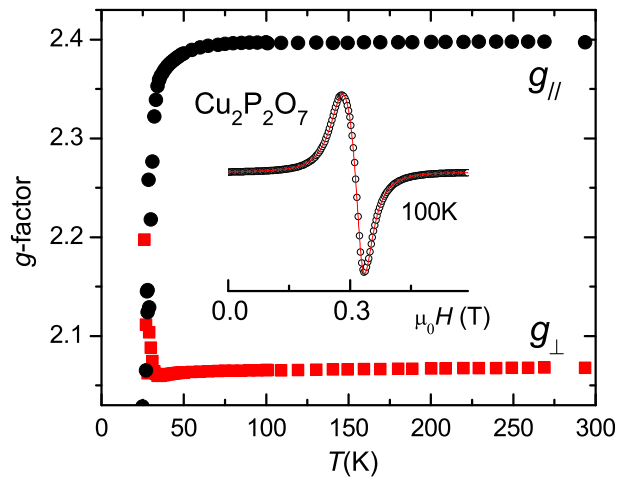}
\caption{\label{F_ESR}(Color online) Temperature dependence of the $g$-factor
measured by the ESR of Cu$^{2+}$. The inset shows a typical derivative ESR
spectrum (circles) fitted by two powder-averaged Lorentzian shapes (solid
line).} 
\end{figure}

In order to estimate the $g$-value and prove the AF ordering in Cu$_2$P$_2$O$_7$, we measured the electron spin
resonance (ESR) of the Cu$^{2+}$ spins as a local probe (Fig~\ref{F_ESR}).  The
temperature dependencies of the integrated ESR absorption and $\chi(T)$ agree
with each other (not shown). At elevated temperatures, both components of the
$g$-tensor ($g_\perp$ and $g_\|$) show typical values for the Cu$^{2+}$ ion. An
abrupt change in $g_\perp$ and $g_\|$ around $T_N$\,=\,27\,K indicates the onset
of a strong internal magnetic field and thus confirms the magnetic ordering
transition observed in the $\chi(T)$ data.

The magnetization $M$($H$) curve (Fig.~\ref{F_exp}, bottom) was scaled using 
the low-field (MPMS) data. It lacks any signatures of a spin gap and exhibits
a linear behavior, incompatible with a simple dimer magnetism.

\subsection{\label{DFT}DFT calculations}
\begin{figure}[tbp]
\includegraphics[angle=270,width=8.6cm]{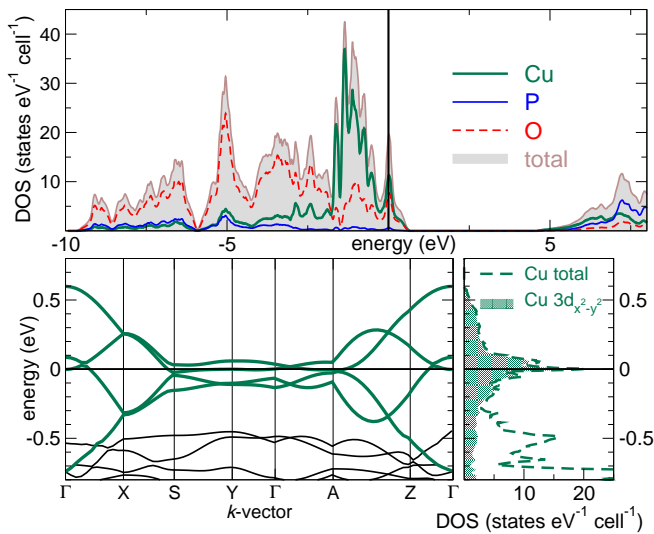}
\caption{\label{F_dosband}(Color online) Density of states (top),
band structure (bottom, left), and orbital-resolved density of states
(bottom, right). The bands with $3d_{x^2-y^2}$ character are shown
bold (green). The Fermi level is at zero energy.}
\end{figure}

To evaluate the leading exchange couplings, we perform DFT band structure
calculations.  For \cupo, LDA yields a valence band with a width of
10\,eV (Fig.~\ref{F_dosband}, top), slightly larger than typical values
for cuprates (7--8\,eV).\cite{kapel_hayd_DFT,Bi2CuO4_DFT} The density of
states (DOS) exhibits a peak at the Fermi level $\varepsilon_{\text{F}}$,
evidencing a metallic GS, in contrast to the expected insulating
behavior. This drawback of the LDA originates from the underestimation
of correlation effects, intrinsic for the $3d^9$ electronic
configuration of Cu$^{2+}$.  To account for the correlation effects, we
use two complementary approaches: (i) mapping the LDA bands onto a
Hubbard model and (ii) adding correlations in a mean-field way within
the band structure scheme (LSDA+$U$ calculations).

Despite the problem of correlations, LDA yields reliable information on
the relevant orbitals and couplings. First, we project the DOS onto
local orbitals of a CuO$_4$ plaquette. We find that the Cu
states at $\varepsilon_{\text{F}}$ have $3d_{x^2-y^2}$ character
(Fig.~\ref{F_dosband}, bottom right), resolving it as the
magnetically active orbital.\footnote{Despite the geometrical distortion of
CuO$_4$ units, resulting in a small hybridization of Cu $3d_{x^2-y^2}$ states
with other Cu states (Fig.~\ref{F_dosband}, bottom left), this hybridization is
far too small to affect the orbital ground state.} Further on, the relevant couplings
were identified using Wannier functions (WF) for the Cu $3d_{x^2-y^2}$
states.\cite{*[{The localization procedure is described in }] [{}] FPLO_WF} The
resulting fit is in excellent agreement with the LDA bands
(Fig.~\ref{F_dosband}, bottom left) and yields five relevant couplings. Two of
them --- the couplings within the structural dimer $t_1$ and between the
structural dimers $t_1'$ (Fig.~\ref{F_str}, middle) --- form alternating chains
along the $b$ axis. The chains are connected by two types of interchain
couplings: $t_{\text{ic}1}$ and $t_{\text{ic}3}$ run through a single
$A$O$_4$ tetrahedron, whereas $t_{\text{ic}2}$ connects two Cu atoms with a double
bridge of $A$O$_4$ tetrahedra (Fig.~\ref{F_str}, bottom).
Together with $t_1$ and $t_1'$, these interchain couplings form the magnetic
layers, parallel to (101), as shown in Fig.~\ref{F_str}. The interlayer
coupling of 30\,meV is realized via pyrophosphate groups P$_2$O$_7$.
Since it is substantially smaller than the leading terms, we neglect
this coupling in the minimal model.

\begin{figure*}[tbp]
\includegraphics[width=17.2cm]{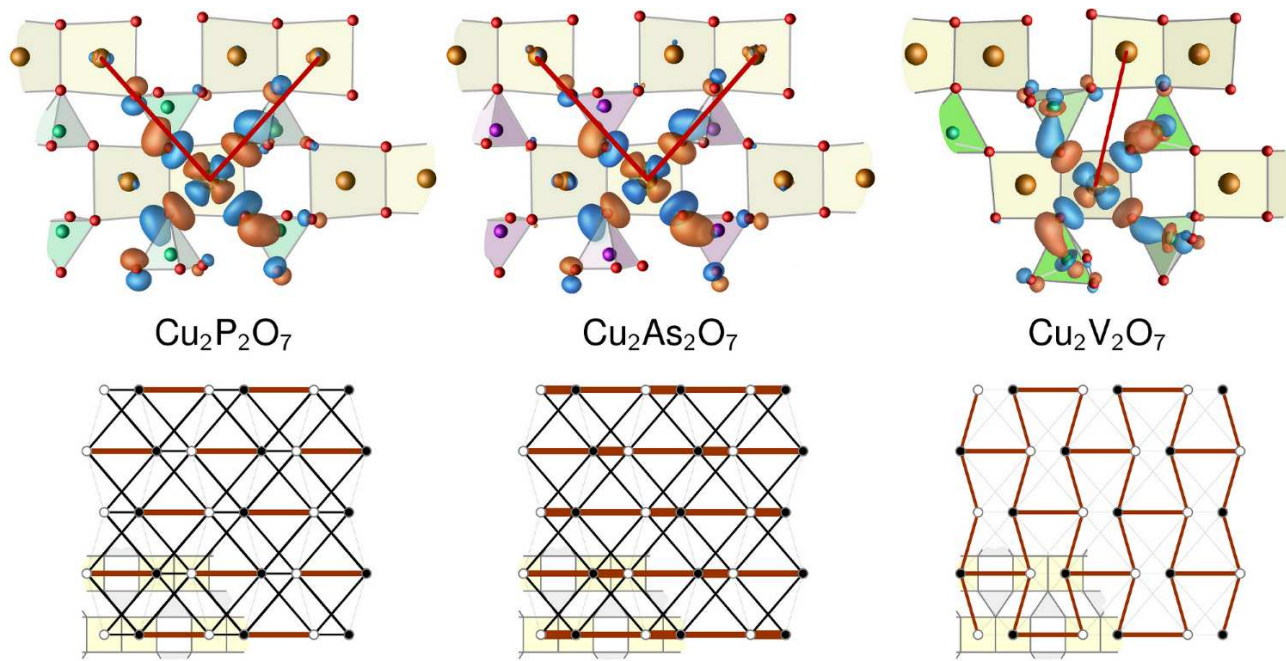}
\caption{\label{F_WF}(Color online) Top: Wannier functions for
Cu $3d_{x^2-y^2}$ states with relevant interchain couplings (lines). Bottom:
spin models for Cu$_2A_2$O$_7$ compounds. The thickness of a line reflects the
magnitude of the respective coupling. Filled and empty circles represent up and down spins in the antiferromagnetically ordered state.} \end{figure*}

Here, we briefly remind the reader of the essential results for
\cuvo\ in Ref.~\onlinecite{HCL_Cu2V2O7_DFT_DCC_simul}: the WF
analysis of the LDA bands revealed that $t_1$, $t_1'$, and $t_{\text{ic}2}$
couplings are relevant, while other terms are negligibly small
(Table~\ref{T_tJ}). A mere comparison of $t_i$ for \cupo~with
the respective values for \cuvo\ readily yields several important
results. First, the coupling $t_1$ within the structural dimer is
essentially the same, while the interdimer coupling $t_1'$ is
slightly enhanced in \cupo~compared to \cuvo. The similarity of the
intra-chain terms is in sharp contrast to the profound change in the
interchain coupling regime: \cupo~has sizable $t_{\text{ic}1}$ and
$t_{\text{ic}3}$ with the negligibly small $t_{\text{ic}2}$, while
for \cuvo~the situation is exactly the opposite. This results in
clearly different band dispersions in \cupo~(Fig.~\ref{F_dosband})
and \cuvo~(Fig.~3 in Ref.~\onlinecite{HCL_Cu2V2O7_DFT_DCC_simul}).

\begin{table}[tbp]
\begin{ruledtabular}
\caption{\label{T_tJ} Transfer ($t_i$, in meV) and exchange ($J_i$, in K)
integrals for \cupo~and \cuvo.  $J^{\text{AF}}_i$ (in K) are calculated
as $4t^2_i/U_{\text{eff}}$ (see text).}
\begin{tabular}{l r r r p{0.8cm} r r r}
path & \multicolumn{3}{c}{\cupo} &
\multicolumn{4}{r}{\cuvo~(Ref.~\onlinecite{HCL_Cu2V2O7_DFT_DCC_simul})}
\\ & $t_i$ & $J^{\text{AF}}_i$ & $J_i$ & & $t_i$ & $J^{\text{AF}}_i$ & $J_i$ \\
$X_1$  &              156 & 251 &  34 & &   148 & 226 & 5 \\ 
$X_1'$ &              103 & 109 & 102 & & $-$84 &  73 & 61 \\
$X_{\text{ic}_1}$ &    83 &  71 &  46 & &    18 &   3 & --\\
$X_{\text{ic}_2}$ & $-$12 &   1 &  -- & &    97 &  97 &  87 \\
$X_{\text{ic}_3}$ &    79 &  64 &  41 & & $-$15 &   2 & --\\
\end{tabular}
\end{ruledtabular}
\end{table}

To obtain the values of AF exchange integrals $J^{\text{AF}}_i$, we map
the transfer integrals $t_i$ from the WF-based analysis onto a Hubbard
model. For the half-filling and the strongly correlated limit
$U_{\text{eff}}\,{\gg}\,t_i$, both well justified for undoped cuprates,
magnetic excitations can be described by a Heisenberg model, while
$J^{\text{AF}}_i$ is expressed as
$J^{\text{AF}}_i$\,=\,$4t^2_i/U_{\text{eff}}$. This way, using a realistic
value of 4.5 eV for the Coulomb repulsion
$U_{\text{eff}}$,\cite{Bi2CuO4_DFT} we obtain the AF part
$J^{\text{AF}}_i$ of the total exchange $J_i$
(Table~\ref{T_tJ}).\footnote{Exchange integrals $J^{\text{AF}}_i$
sharpen the trends inferred from the analysis of $t_i$'s:
$J^{\text{AF}}_{\text{ic}_1}$ and $J^{\text{AF}}_{\text{ic}_3}$ are
relevant for \cupo, while $J^{\text{AF}}_{\text{ic}_2}$ is relevant for
\cuvo.}

For the NN exchange coupling path $X_1$, the Cu--O--Cu angle is close to
90$^{\circ}$ (98.7$^{\circ}$ for \cuvo, 100.4$^{\circ}$ for \cupo),
suggesting a sizable ferromagnetic contribution according to the GKA
rules.\cite{GKA_1, *GKA_2, *GKA_3} Indeed, in many related systems this
leads to a sizable reduction in the respective exchange
$J_i$.\cite{*[{See }] [{ for an instructive example.}]
FLD_BiCu2PO6_DFT_simul_chiT_MH} Moreover, in case the FM contribution
exceeds $J^{\text{AF}}_i$, the resulting $J_i$ exchange becomes
FM.\cite{HC_CuSiO3H2O_DFT_QMC_chiT} Therefore, to complete the spin
model, it is necessary to calculate the values of total exchange
integrals $J_i$, being the sum of the AF and FM contributions. We
evaluate the numerical values by performing spin-polarized LSDA+$U$
supercell calculations for various collinear spin arrangements and map the
total energies onto a classical Heisenberg model.  Following the conclusions of
Ref.~\onlinecite{HCL_Cu2V2O7_DFT_DCC_simul}, we choose the around-mean-field
double-counting-correction (DCC) and $U_d$\,=\,6.5\,eV.\footnote{The access to all
relevant couplings is ensured by considering three different supercells.} The
resulting exchange integrals $J_i$ are given in
Table~\ref{T_tJ}.\footnote{LSDA+$U$ calculations correctly reproduce the
insulating ground state in the physically relevant range of
$U_d$\,=\,$5.5-7.5$\,eV, yielding the band gap of about 3.3(6)\,eV for
Cu$_2$P$_2$O$_7$ and 2.9(3)\,eV for Cu$_2$As$_2$O$_7$.}

For \cupo, a sizable FM contribution of $J^{\text{FM}}_1\sim-230$~K
considerably reduces the coupling within the structural dimer, as in \cuvo.
Again similar to \cuvo, the interdimer coupling $J_1'$ has a tiny FM
contribution only.\cite{*[{Parallel alignment of two well-separated CuO$_4$
plaquettes strongly reduces the overlap of O orbitals belonging the different
plaquettes. Therefore, the ferromagnetic contribution becomes vanishingly small. This has
been proven for several systems, for instance Sr$_2$Cu(PO$_4$)$_2$, discussed
in }] [{}] HC_Sr2CuPO42_DFT_chiT_fit} However, in \cupo, $J_1'$ is clearly
dominant and seemingly favors a dimer-like magnetism with magnetic dimers
located between the structural dimers Cu$_2$O$_6$. All interchain
couplings are long-range, and their FM contributions are likely to be
small. Indeed, we find
$J^{\text{FM}}_{\text{ic}_1}\,{\approx}\,J^{\text{FM}}_{\text{ic}_3}\,{\sim}\,-$20\,K
consistent with $J^{\text{FM}}_{\text{ic}_2}$\,=\,$-10$\,K found for
\cuvo.\cite{HCL_Cu2V2O7_DFT_DCC_simul} Noteworthy, for the closely related
system Cu$_2$(PO$_3$)$_2$(CH$_2$), similar values have been
reported.\cite{Cu2PO32CH2_DFT_NMR_chiT_CpT_MH_simul}  To summarize, for
\cupo~we arrive at the two-dimensional (2D) magnetic model with a
strongly dominant $J_1'$ and three additional relevant couplings,
rather different from the 2D honeycomb lattice system \cuvo\ as
well as from the structurally-related but essentially one-dimensional
(gapped $J_1-J_1'$ chains) (VO)$_2$P$_2$O$_7$ in its high-pressure
modification.\cite{[{}][{, and references therein.}]vpo}

Despite the 2D nature of the spin models, the magnetic GSs of \cupo\ and
\cuvo\ are AF-ordered.\cite{neutrons,cuvo_ord} Since thermal 
fluctuations in purely 2D Heisenberg systems exclude long-range magnetic
order at finite temperatures,\cite{Mermin_Wagner} the coupling
perpendicular to the magnetic layers should be addressed. In \cuvo,
effective superexchange paths along the $[$V$_2$O$_7]$ pyrovanadate
groups give rise to a sizable long-range interlayer exchange
$J_{\perp}$\,=\,17\,K.\cite{HCL_Cu2V2O7_DFT_DCC_simul} In contrast, the
interlayer couplings in \cupo\ are much weaker and short-range: although
the transfer integral associated with the shortest interlayer coupling
(Cu--Cu distance 3.33\,\AA) amounts to $t_{\perp1}$\,=\,31\,meV, the LSDA+$U$
calculations reveal a sizable FM contribution that largely reduces this
coupling to about $J_{\perp1}$\,=\,1\,K.  The next largest
($t_{\perp2}$\,=\,18\,meV) interlayer coupling corresponds to the Cu--Cu
distance of 3.88\,\AA, and gives rise to a magnetic exchange of 3--4\,K. 

Using the interlayer couplings resulting from the DFT calculations, we find one more
difference between \cupo\ and \cuvo. In \cuvo, the magnetic structure
should break the $C$-centering of the unit
cell,\cite{HCL_Cu2V2O7_DFT_DCC_simul} whereas the long-range order in
\cupo\ inherits the full crystallographic symmetry (the $C2/c$ space
group). The distinct patterns of the magnetic ordering should be resolvable by
high-precision neutron diffraction and could be an ultimate experimental
test of our predictions. The lack of magnetic
reflections in the available neutron data for \cupo\
(Ref.~\onlinecite{neutrons}) indicates that the magnetic order
does not break any crystallographic symmetry, nor leads to the formation of a
magnetic superstructure, in accord with our microscopic model.

\subsection{\label{mdl_sml}Model simulations}
To justify the proposed spin model for \cupo, we perform QMC simulations
of the magnetic susceptibility and the high-field magnetization data. First, to fit
the calculated susceptibility to the experiment (Fig.~\ref{F_exp}), we
keep the ratios $J_1/J_1'$\,=\,0.33, $J_{\text{ic}_1}/J_1'$\,=\,0.45, and
$J_{\text{ic}_3}/J_1'$\,=\,0.40 from the DFT calculations
(Table~\ref{T_tJ}), and vary the absolute values of $J_1'$ and
$g$.\footnote{The details of the fitting procedure are extensively
discussed in Ref.~\onlinecite{HC_CuSiO3H2O_DFT_QMC_chiT}} This way, we
obtain an excellent agreement with the experimental curve down to the
ordering temperature $T_N$\,=\,27\,K (Fig.~\ref{F_exp}). The resulting
$J_1'$\,=\,79\,K and $g$\,=\,2.21 are in reasonable agreement with $J_1'$\,=\,102\,K from
the LSDA+$U$ calculations and the powder-averaged
$g$\,=\,$(2g_{\perp} + g_{\parallel})/3$\,=\,2.18 from
ESR.

The simulated magnetization curve (Fig.~\ref{F_exp}, bottom) is scaled
adopting $J_1'$\,=\,79\,K and $g$\,=\,2.21 from the fit to the magnetic
susceptibility. The overall agreement with the experiment is rather
good, while the discrepancy likely arises from the imperfectness of the scaling
of the experimental high-field curve. In particular, the scaling is done based
on a very narrow low-field region (up to 5\,T, see the inset of
Fig.~\ref{F_exp}), which might be affected by paramagnetic
impurities.\footnote{The uncertainty of such scaling can reach 30--40\,\%,
which largely exceeds typical uncertainties of $\chi(T)$ measurements.
Therefore, the fitting of the magnetic exchange couplings is done using
$\chi(T)$ data only.}

\section{\label{disc}Discussion}
The spin models of \cupo~and \cuvo~are essentially different:
whereas \cupo~is a system of coupled dimers, \cuvo~is an almost
isotropic honeycomb lattice. In \cuvo, a strong FM
contribution practically ``switches off'' the coupling within the
structural dimers ($J_1$), leaving only the couplings between the
structural dimers ($J_1'$) and between the chains ($J_{\text{ic}_2}$)
relevant (Fig.~\ref{F_WF}, bottom
right).\cite{HCL_Cu2V2O7_DFT_DCC_simul} On the contrary, \cupo~has a
dominant $J_1'$, smaller $J_{\text{ic}_1}$ and $J_{\text{ic}_3}$
(see Fig.~\ref{F_str} for the notation of the exchange couplings) as
well as even smaller $J_1$ (Fig.~\ref{F_WF}, bottom left). The
striking dissimilarity of the two models leads to a question, whence
this difference originates.

To address this puzzling issue, we consider another \cua~system,
\cuas, and briefly report the key results of our DFT calculations.
In particular, the WF analysis yields values of transfer integrals
$t_i$ similar to those of \cupo. The total exchange integrals,
derived from the results of LSDA+$U$ calculations, are somewhat
larger than the respective exchange couplings for \cupo.  However,
the only pronounced difference between the magnetic models of
\cupo~and \cuas~is the strong enhancement of the NN coupling $J_1$
in the latter system (Fig.~\ref{F_WF}, bottom middle). A detailed
experimental and theoretical analysis of the magnetic properties of \cuas~is
presently underway.\cite{Cu2As2O7_DFT_chiT_CpT_ESR_NMR}

So far, we find the resemblance of the spin models of \cupo~and \cuas,
whereas \cuvo~is remarkably different. The common feature of these
compounds is the sizable AF coupling $J_1'$ forming spin dimers. The
interdimer couplings are, however, different. In \cupo~and \cuas, the
interdimer couplings run through a single $A$O$_4$ tetrahedron
($J_{\text{ic}_1}$ and $J_{\text{ic}_3}$), while in \cuvo~only the
coupling via a double bridge of AO$_4$ tetrahedra is relevant.

As we know from the analysis of the band structure, the magnetism of
\cua~systems originates from the half-filled Cu $3d_{x^2-y^2}$
orbitals.  Above, we used the WF for this magnetically active
orbital to obtain the relevant couplings (Sec.~\ref{DFT}). However,
the fact that WF are defined in real space, makes them capable of
tracing the superexchange path.\cite{*[{For example, see Fig.~6
in }] [{}] HC_CuSe2O5_DFT_chiT_CpT_simul}
In Fig.~\ref{F_WF}, we plot the WF for the three systems.

As expected from the similarity of the band structures, the WFs for
\cupo~and \cuas~are very similar.  \footnote{The only visible
difference is the larger weight of the Wannier function on the neighboring copper
atom (corresponding to $t_1$ coupling) in the case of \cuas, in line
with the enhanced value $J_1$ in \cuas~compared to \cupo.} Note (i)
the absence of P or As contributions and (ii) the antibonding
combination formed by WF contributions along Cu--O--O--Cu paths, favoring
$t_{\text{ic}1}$ and $t_{\text{ic}3}$ couplings. In contrast, the WF of
\cuvo~shows sizable weight on V atoms, exhibiting a shape reminiscent of the
$d_{3z^2-r^2}$ orbital. These V states hybridize with the nearby O orbitals,
thus forming an effective superexchange path Cu--O--V--O--Cu.  The V--O
hybridization drastically alters the interchain coupling regime, now favoring
$J_{\text{ic}_2}$ exchange via a double bridge of VO$_4$ tetrahedra.

As evidenced by numerous examples, the replacement of anionic groups can
stabilize different crystal structures. Thus, the replacement of Se by a larger
Te atom transforms the chain-like structure of CuSe$_2$O$_5$ into complex 2D
layers of CuTe$_2$O$_5$. Not surprisingly, this structural change drastically
affects the magnetism: while CuSe$_2$O$_5$ shows Heisenberg chain physics and
AF order,\cite{HCL_Cu2V2O7_DFT_DCC_simul} CuTe$_2$O$_5$ has a dimerized
magnetic GS.\cite{CuTe2O5_chiT_ESR_EHTB} However, it is commonly believed that
the replacement of a nonmagnetic part, \emph{not~accompanied} by structural
changes, should have a minor impact on magnetism. Our study shows that this
conjecture is not universal. In particular, the electronic configuration of the
nonmagnetic group should be addressed. Thus, the $d$-states of V in \cuvo~are
strongly mixed with the $p$ states of oxygen (see Fig.~2 in
Ref.~\onlinecite{HCL_Cu2V2O7_DFT_DCC_simul}), allowing for a strong
superexchange along the largely covalent V--O bonds. On the contrary, $p$-cations
like P or As donate electrons to the neighboring O atoms, giving rise to more
ionic P(As)--O bonds. In this case, electron hopping between P(As) and O atoms
becomes energetically unfavorable, thus confining the superexchange to the
Cu--O--O--Cu path.

Regarding the superexchange in transition-metal phosphates, the lack of
the P contribution to the magnetic orbital and the crucial role of 
$M$--O--O--$M$ pathways has been recently established for the related
spin-$\frac12$ V$^{+4}$ compounds. A systematic study of model
structures with different bond
distances and angles puts forward the $M$--O and O--O distances as key
structural parameters that determine the coupling
strength.\cite{sqlat,pbznvp2o9} Another relevant feature is the number
of bridging tetrahedral groups. Given similar bond distances, a double
bridge makes two superexchange pathways, which are more efficient than
a single bridge.\cite{roca1998} This trend holds for \cupo, where
the double-bridge coupling $J_1'$ is stronger than the single-bridge
exchanges $J_{\text{ic}1}$ and $J_{\text{ic}3}$ (Table~\ref{T_tJ}). An
important difference between copper and vanadium phosphates is the
stronger metal--oxygen hybridization in the former case. The
hybridization, amplified by the $\sigma$-type bonding between Cu
$d_{x^2-y^2}$ and oxygen $p$ orbitals, leads to a non-negligible
contribution of second-neighbor oxygen atoms to WF (see
Fig.~\ref{F_WF}). In BiCu$_2$PO$_6$, such ``tails'' of WF are
responsible for the pronounced difference between the seemingly
equivalent long-range couplings.\cite{FLD_BiCu2PO6_DFT_simul_chiT_MH} A
comparative study of Cu$^{+2}$ and V$^{+4}$ compounds could facilitate
further development of empirical rules for the superexchange. The \cua\
structures are especially apposite for this purpose due to their
similarity to (VO)$_2$P$_2$O$_7$, CsV$_2$O$_5$, VOSeO$_3$, and other
V$^{+4}$ oxides.

Generalising the results, our study evidences that for the tetrahedral
non-magnetic $A$O$_4$ groups with the central $p$-atom the magnetic
superexchange is realized via an O--O edge of the tetrahedron. In contrast, a
tetrahedrally coordinated $3d$ element gives rise to a strong
hybridization with O $2p$ orbitals and a concomitant change in the
superexchange path.  This finding constitutes an empirical rule to
estimate the relevant long-range couplings in a wide range of
low-dimensional systems containing non-magnetic tetrahedral side groups.

\section{\label{summ}Summary and outlook}
To summarize, we presented a comprehensive experimental (magnetic
susceptibility, ESR, and high-field magnetization) and theoretical
(full-potential DFT calculations) investigation of the magnetism of
\cupo. This material implies a two-dimensional magnetic model of
strongly coupled magnetic dimers, which interestingly does not lead
to a singlet ground state: sizable interdimer couplings enforce the
antiferromagnetic ordering at 27\,K.

The coupled dimer model of \cupo~is essentially different from the
honeycomb lattice model, realized in the isostructural system \cuvo.
Based on a comparison between \cupo, \cuas, and \cuvo, we unravel
the crucial importance of the non-magnetic $A$O$_4$ side groups for
the magnetism of \cuaf~systems: whereas a $d$-atom (V) strongly
hybridizes with the ligand O orbitals, the $p$-states (P, As) do not
contribute to the superexchange path. Within the structure of
\cua~compounds, this leads to a drastic change in the topology of
the interchain couplings. The forthcoming experimental study of \cuas~should provide a
deeper insight into the magnetism of this system and challenge our
findings. Our results can be applied to tune the magnetism of
systems comprising a similar structural motive: for instance,
Cu$_2$(PO$_3$)$_2$(CH$_2$)
(Ref.~\onlinecite{Cu2PO32CH2_DFT_NMR_chiT_CpT_MH_simul}) or
(VO)$_2$P$_2$O$_7$ (Ref.~\onlinecite{vpo}).

Our study emphasizes the danger to use isostructural compounds as
reference or analog. Such comparisons, unless based on a precise
knowledge of the underlying electronic structure, are at best not
justified and can be completely misleading. Instead, they are amenable
to a careful verification. The latter can be reliably performed by
electronic structure calculations that provide a microscopic insight
into the electronic and magnetic properties even of complex compounds.

\section{Acknowledgements}
We are grateful to Yurii Prots and Horst Borrmann for X-ray diffraction
measurements, and to Walter Schnelle for fruitful discussions and
valuable comments. Part of this work has been supported by EuroMagNET II
under EU Contract No. 228043. A.~T. acknowledges funding from
Alexander von Humboldt Foundation.

%

\end{document}